\documentstyle[preprint,aps,eqsecnum]{revtex}
\begin{document}
\title
{Ground state energy and
quasiparticle  gaps\\
 in $\nu={N\over{2N\pm 1}}$ FQHE states}

\author{D.V.Khveshchenko}
\address
{Physics Department, Princeton University,\\ Princeton, NJ 08544\\
and\\
Landau Institute for Theoretical Physics,\\
2,st.Kosygina, 117940, Moscow, Russia}
\maketitle

\begin{abstract}
\noindent
Applying the transformation of fermion operators to new fermion
quasiparticles  introduced by Halperin, Lee, and Read we estimate
a scaling behavior of the ground state energy and quasiparticle
gaps as a function of filling fraction for a "principal sequence"
of FQHE $\nu={N\over{2N\pm 1}}$ states converging  towards the gapless
state at half filling. The exponent describing the shape of
the cusp $\delta E(\nu)\sim |\delta\nu|^{\eta}$
 is found to be greater than one and to depend nontrivially
on the interaction potential. The dependence of quasiparticle gaps
agrees with the results of recent measurements by R.R.Du et al.
\end{abstract}
\pagebreak

The Fractional Quantum Hall Effect still remains to be among the most
intensively studied subjects of the modern Condensed Matter Theory [1].
More generally, it forms part of the problem of the global  phase
diagram of interacting two-dimensional fermions in an external
magnetic field.

Methodologically, understanding this seemingly
 special issue could, in principal,
be  crucially important in view of novel scenarios
of spontaneous parity- and time-reversal symmetry violation due to
strong interactions [2]. The parity-odd ground state resulting from
 such an instability is supposed
to be characterised by a spontaneously generated flux of some fictitious
magnetic field associated with a phase of the bi-local operator\\
 $\Delta(x,x^{\prime})=\Psi^{\dagger}_{\sigma}(x)\Psi_{\sigma}(x^{\prime})$
 [3].

Recently the theory of FQHE received a novel trend of development with
 a new look at the nature of states with even denominators [4],
and in
particular,  the $\nu={1\over 2}$ state which is of most
experimental interest. The results of measurements of
resistivity and surface acoustic wave attenuation indicate a presence of a
Fermi
surface (in the Luttinger sense) consistent with the density of carriers [5].

Because of the intrinsic degeneracy of a partially filled Landau level
any analytical calculations in the framework of FQHE are quite complicated.
Although numerous symmetry considerations,
variational techniques and various phenomenological
approaches (see [1] for references)
provided an essential progress in understanding
of the FQHE phenomena, a consistent  method of microscopic calculations
is still strongly required. Basically, most of present knowledge about
the FQHE concerns those universal properties (such as a quantization
of the Hall conductivity, quantum numbers and statistics of
quasiparticles) which are independent, to a significant extent,
of the details of the
interaction. However the latter become very important if one
intends to estimate those
experimentally measurable characteristics as quasiparticle
gaps or to establish the dependence of the ground state energy $E(\nu)$
on filling fraction $\nu$.

 In Ref.[4] a new formalism to treat FQHE
fractions with even denominators was proposed.
The basic idea first discussed by Jain [6]
is to attach to each fermion an even number of fluxes of
some fictitious ("statistical") field. Under this transformation fermions
preserve their statistics, but at the mean field level
an external magnetic field applied to fermions
can be cancelled by an average fictitious one.
One then obtains fermions in a zero net field interacting via fluctuations
of the fictitious flux.

This procedure can be understood as a particular
way to construct
 "proper zero order eigenstates" which lifts up the Landau level
degeneracy as a result of the
 electron interaction $V(r)={e^{2}\over{r^{\alpha}}}$.

In what follows we shall restrict  ourselves to the even-denominator
case $\nu={1\over 2}$.
It was also proposed in [4] to treat odd-denominator
"descendent" $\nu={N\over{2N\pm 1}}$
FQHE states within the same framework. Namely, one can keep attaching
two flux quanta to each particle, the net field being equal to
$B={2\pi\rho\over N}$,
where $\rho$ is a fermion density. Then to find the effect of
the original interaction $V(r)$ one has to consider a situation of
the IQHE with additional interactions due to fluctuations
of the fictitious flux.

In the present paper we  study a dependence of the ground state energy
on $N$ in the vicinity of
half filling $(N>>1)$. More concretely, we intend to find the
 shape of the cusp
 $\delta E(\nu)\sim |\nu-\nu_{c}|^{\eta}$ at $\nu_{c}={1\over 2}$.
We also estimate a scaling behavior of the excitation gaps.

It is widely believed that the function $E(\nu)$ is not differentiable
at all rational points and presents an example of a fractal. Evidently,
this circumstance makes it impossible to get any analytic formula
describing $E(\nu)$ at all values of $\nu$. However one can find the
 behavior of
$E(\nu)$ on some restricted support, for instance, on a sequence of values
$\nu={N\over{2N\pm 1}}$ and to determine an exponent $\eta$
for this case. It is quite possible, of course, that choosing
another sequence one will obtain another value of $\eta$. Nevertheless,
from the experimental point of view a determination of $\eta$ for a given
sequence of FQHE states still has sense, because not all cusps are equally
prominent, hence  not all sequences are
equally well observed.

We shall demonstrate an importance of fluctuations
of the fictitious
flux which change
the dependence of
the ground state energy
on the form of
interaction potential with
respect to the result of the naive
Hartree-Fock approximation.
However, at least at weak couplings these strong fluctuations
don't make the result universal yet.
In particular, we obtain that $\eta$  depends on the exponent $\alpha$
governing
a power-like decay of the bare potential $V(r)$, although this dependence
is quite nontrivial.

To proceed with calculations we use the conventional Lagrangian
\begin{equation}
L=\int d^{2}r
\Psi^{\dagger}(i{\partial\over{\partial t}}-
{1\over{2m}}(-i{\vec \nabla}-{\vec A})^{2}+\mu)\Psi
+{1\over 2}\int d^{2}r\int d^{2}r^{\prime}
\Psi^{\dagger}\Psi(r) V(r-r^{\prime})
\Psi^{\dagger}\Psi(r^{\prime})
\end{equation}
where ${\vec A}$  denotes an external field and a chemical potential
$\mu$ corresponds to $\nu={N\over{2N\pm 1}}$.

By introducing two flux quanta per fermion by means of the well-known
"Chern-Simons transformation" one obtains an equivalent Lagrangian
in terms of an additional gauge field $(\Phi,{\vec a})$ [4]:
\begin{eqnarray}
L=\int d^{2}r
\Psi^{\dagger}(i{\partial\over{\partial t}}-
{1\over{2m}}(-i{\vec \nabla}-{\vec A}-{\vec a})^{2}+\mu)\Psi\nonumber\\
+{1\over{32\pi^2}}\int d^{2}r\int d^{2}r^{\prime}
{\vec \nabla}\times{\vec a}(r)V(r-r^{\prime}){\vec \nabla}\times{\vec a}
(r^{\prime})
+\Phi(\Psi^{\dagger}\Psi -{1\over 4\pi}{\vec \nabla}\times{\vec a})
\end{eqnarray}
where we identified density and "statistical" (gauge) flux fluctuations
($\delta\rho={1\over 4\pi}{\vec \nabla}\times \delta{\vec a}$)
due to the local constraint ${\partial L\over {\partial\Phi}}=0$.
Then one can consider a mean field solution
characterised  by a uniform total field $B={2\pi\rho\over N}$.
 In fact, the stability of this
mean field solution might be crucially dependent on the  strength
of the interaction potential $e$.
It is unclear now whether such a solution becomes stable at some
threshold value of $e$ or arbitrarily small $e$ are also
acceptable. In what follows we assume
 that the threshold value  (if any) is small enough compared with
$1/N$.

In the absence of interactions the mean field energy at all $N$
is, of course, equal to the kinetic energy of free fermions $E_{0}(\nu)
={\pi\rho^{2}\over m}$
(we remind the
reader that the energy of 2D fermions occupying an integer number of
Landau levels is degenerate with the energy of free fermions with a circular
Fermi surface).

 Formally putting $e=0$ we have to recover the energy of a partially filled
 lowest Landau level $E_{0}(\nu)
={\pi\rho^{2}\over{\nu m}}$. It can be immediately seen that the role of
 gauge fluctuations is quite important. In particular, in the
mean field picture one-particle excitations have a pseudogap
$\Delta={B\over m}$
which survives in the
absence of any real interaction. This artifact has to disappear, of course,
after a proper account of gauge fluctuations.

Although a
 complete account of those fluctuations doesn't seem possible,
one could assume that to analyze the effects of the interaction $V(r)$
it suffices to take into account the leading contributions corresponding to
the conventional RPA. This approximation is supposed to be adequate
at least in the vicinity of the half filled case (that is, at $N>>1$).
We shall also comment on this point below.

In RPA one encounters the  problem of calculating the determinant
of a quadratic operator
which governs gaussian fluctuations of the gauge field
\begin{equation}
K_{\mu\nu}(\omega, k)=\Pi_{\mu\nu}(\omega, k)
+{1\over 4\pi}(\delta_{\mu i}\delta_{\nu 0}+
\delta_{\mu 0}\delta_{\nu i})\epsilon_{ij}k_{j}+
\delta_{\mu i}\delta_{\nu j}
{1\over (4\pi)^2}V(k)(\delta_{ij}k^{2}-k_{i}k_{j})
\end{equation}
where $\Pi_{\mu\nu}(\omega, k)=<J_{\mu}(-\omega, -k)J_{\nu}(\omega, k)>
$ denotes a fermion polarization operator in a uniform magnetic field
$B={2\pi\rho\over N}$ and
 $V(k)\sim e^{2}k^{\alpha-2}$
stands for a Fourier transform of the interaction
potential (in the case of
the short-range interaction $V(r)\sim \delta(r)$ we put $\alpha=2$).

The $a ~priori$ knowledge of the effects of the gauge
interaction at $e=0$
allows one to use the following normalization of the RPA  energy
correction
\begin{equation}
E(\nu)=\int {d\omega\over{2\pi}}
\int {d^{2}k\over{(2\pi)^{2}}} Im (\ln Det {\hat K}-\ln Det {\hat K_{0}})
\end{equation}
Expanding (4) as a series in $e^{2}$ we find the lowest order term in the form
\begin{eqnarray}
E_{1}(\nu)=-{1\over (4\pi)^2}\int {d\omega\over{2\pi}}
\int {d^{2}k\over{(2\pi)^{2}}}
Im{\Pi_{00}V(k)k^{2}\over
{Det {\hat K}_{0}}}=\nonumber\\
= -{1\over (4\pi)^2}
\int {d\omega\over{2\pi}}
\int {d^{2}k\over{(2\pi)^{2}}}V(k)k^{2}Im <a_{\perp}(\omega, k)
a_{\perp}(-\omega, -k)>
\end{eqnarray}
where we introduced a propagator of the transverse
component of the gauge field $(a_{\perp}={{\vec a}\times{\vec k}\over k})$.
Notice, in passing, that the absence of poles in the propagator
$<a_{\perp}(\omega, k)a_{\perp}(-\omega, -k)>$ at $\omega <{B\over m}$ and $k<
{B\over \sqrt\rho}$
follows from the incompressibility of the mean field state.

To facilitate the calculation of (5) one has to find a treatable
representation for the polarization operator  $\Pi_{\mu\nu}(\omega,k)$.
This subtle problem is a long-standing one. At arbitrary number of
occupied Landau levels $N$ one has the
following exact formulae $(\omega\neq 0)$:
\begin{equation}
\Pi_{\mu\nu}(\omega, k)={B^{2}m\over{(2\pi)^{2}}}
\sum_{l=N}^{\infty}\sum_{n=0}^{N-1}{(l-n)\over{\omega^{2}m^2-B^{2}
(l-n)^{2}+i0}}<n|J_{\mu}(k)|l><l|J_{\nu}(k)|n>+\delta_{\mu i}\delta_{\nu j}
\delta_{ij}{\rho\over m}
\end{equation}
Three independent components of
$\Pi_{\mu\nu}(\omega, k)$
can be expressed in terms of
 matrix elements of the density $J_{0}(k)$ and the transverse
current $J_{\perp}(k)$ given by the formulae
\begin{equation}
<n|J_{0}(k)|l>={1\over m}e^{-x/2}{\sqrt{Bn!\over{l!}}}(x)^{{l-n\over 2}}
L_{n}^{l-n}(x)
\end{equation}
and
\begin{equation}
<n|J_{\perp}(k)|l>={1\over m}e^{-x/2}{\sqrt{Bn!\over{l!}}}(x)^{{l-n\over 2}}
((l-n-x)L_{n}^{l-n}(x)+2x{d L_{n}^{l-n}(x)\over{dx}})
\end{equation}
where $x= {k^{2}\over 2B}$ and $L_{n}^{s}(x)$ is a Laguerre polynomial.
One can easily find the determinant standing at (5) in
terms of these components
 $Det {\hat K}_{0}=({k\over 4\pi}-\Pi_{odd})^{2}-
\Pi_{00}\Pi_{\perp}$.

Notice that to estimate (5) one has to know the behavior of
$\Pi_{\mu\nu}(\omega, k)$ at
large $\omega$ and $q$.
Because of that the
 expressions (7,8) can be of practical use only at small $N$.
In the opposite limit $N>>1$ it appears to be possible to use
a semiclassical approach elaborated recently by Wiegmann and Larkin [7].
The essence of this method can be demonstrated on an example
of the scalar component of the polarization operator $\Pi_{00}(\omega,k)$.

In the mixed
representation the one-particle Green function can be written in the form
\begin{equation}
G(\epsilon, {\vec r})={B\over{2\pi}}
e^{iBxy-Br^{2}/4}
\sum_{n=0}^{\infty}{L_{n}(Br^{2}/2)\over{\epsilon+\mu-B(n+1/2)+
i\delta}}
\end{equation}
where $\delta = i0 sgn\epsilon$.
In a weak field (9) can be approximated by the expression
\begin{equation}
G(\epsilon, {\vec r})\approx {B\over{2\pi}}e^{iBxy}\sum_{n=0}^{\infty}
{J_{0}({\sqrt{2Bn}}r)\over{\epsilon+\mu-B(n+1/2)+
i\delta}}
\end{equation}
Using (10) one can find the following semiclassical formula
\begin{equation}
\Pi_{00}(\omega, k)\approx {1\over 4\pi}\sum_{n=0}^{N-1}\sum_{s=-n}^{N-1-n}
{sB^{3}m\over{\omega^{2}m^2 -B^{2}s^{2}+i0}}
((2nBk^{2}-(sB-k^{2}/2)^{2})^{-1/2}
\end{equation}
Comparing this expression with the integral representation of the zero field
free fermion polarization
\begin{equation}
\Pi_{00}^{(0)}(\omega, k)=\int^{p_{F}}_{0} {d^{2}p\over (2\pi)^2}
\int^{p_{F}^{2}-p^{2}\over 2}_{k^{2}/2-pk}
 ds
{sm\over{\omega^{2}m^2-s^{2}+i0}}((p^{2}k^{2}-(s-k^{2}/2)^{2})^{-1/2}
\end{equation}
where $p_{F}=\sqrt{4\pi\rho}$,
 we see that (11) can be understood as a result of a "discretization"
of the scattering angle $\theta\rightarrow\cos^{-1}{Bs+k^{2}/2\over {pk}}$
as well as energy levels $p^{2}/2\rightarrow Bn$ in a weak magnetic field.

It can be readily shown that fluctuations of the gauge field
are crucially important for a proper calculation
of even the lowest order correction (5) to the ground state energy.
Indeed, if one completely discards those fluctuations then the lowest order
correction (5) amounts to the exchange energy in a magnetic field
\begin{equation}
E_{ex}(N)=-{1\over 2}\int {d^{2}k\over{(2\pi)^{2}}}V(k)
\int d\omega\Pi_{00}(\omega,k)
\end{equation}
The most convenient way to proceed with (13) is to use a real
 space representation.

By integrating (9) over energy it is easy to show that
an equal time polarization operator can be found in the following
simple form
\begin{equation}
\Pi_{00}(r)=G(r)G(-r)=({B\over{2\pi}})^{2}
e^{-Br^{2}/2}(L^{1}_{N-1}(Br^{2}/2))^{2}
\end{equation}
Although the integral (13) can be found exactly  we shall apply the
asymptotics of Laguerre polynomials in terms of Airy functions to find
the leading term in $1\over N$ expansion
\begin{equation}
e^{-x/2}L_{N}^{1}(x)\approx {1\over{2(2N)^{1/3}}}
Ai((4N)^{2/3}({x\over{4N}}-1))+...
\end{equation}
As a result one arrives at the following formula
\begin{equation}
E_{ex}(N)=
-{8\over {3\pi^2}}e^2 ({p_{F}\over 2})^{2+\alpha}
(1+{c_{0}\over {N^{1+\alpha}}}+...)
\end{equation}
where $c_0$ is positive.
This result is in agreement with a known fact that the exchange energy
is lower in a magnetic field than without it. Notice, in passing, that
this circumstance might play an important role in the context of recent
scenarios of a spontaneous gauge
flux generation due to strong interactions [3].

However the result (16) is in contrast with a general expectation that
the $\nu={1\over 2}$ state is a local minimum. To obtain a correct
sign of the energy correction one has to take into
account gauge fluctuations as well.

A lengthy analysis based on the semiclassical formulae
of the type (11) leads to a remarkably simple prescription.
Namely, it turns out that to estimate the leading term of the $1/N$-expansion
one only needs to know $\Pi_{\mu\nu}(\omega,k)$ at $\omega >{B\over m}$
and $k>{B\over \sqrt\rho}$. Moreover
in this region one can use the  zero field polarization
operator
\begin{eqnarray}
\Pi_{00}\approx -\kappa+i\gamma{\omega\over k}\nonumber\\
\Pi_{\perp}\approx \chi k^{2}+i\gamma{\omega\over k}\nonumber\\
\Pi_{odd}\approx 0
\end{eqnarray}
where $\kappa\sim m$, $\chi\sim m^{-1}$ and $\gamma\sim p_{F}$.
The corrections to the approximate
expressions (17) give  higher order terms in $1/N$.

It can be also shown that the region of small
$x={\omega m\over B}, y= {k\over {\sqrt B}}$,
where the expressions for $\Pi_{\mu\nu}(\omega,k)$ are given by formulae
\begin{eqnarray}
\Pi_{00}\approx -N^{2}m y^{2}(1+N^{2}(-x^{2}+{3\over 8}y^{2}))\nonumber\\
\Pi_{\perp}\approx {N^{2}\over m}(-x^{2}+y^{2})\nonumber\\
\Pi_{odd}\approx {Ny\over 2\pi}(-1+N^{2}(-x^{2}+{3\over 4}y^{2})),
\end{eqnarray}
doesn't contribute to the leading term.
Then the dependence on the field enters only as an effective lower bound
of integrations over $\omega$'s. Performing this approximate calculation
we obtain
\begin{eqnarray}
E_{1}(N)\approx  \int_{B}^{\mu} {d\omega\over {2\pi}}\int_{0}^{\infty}{kdk\over
{2\pi}}
Im{(\kappa-i\gamma\omega/k)V(k)k^{2}
\over{k^{2}+(4\pi)^2(\kappa
-i\gamma\omega/k)(\chi k^2+i\gamma\omega/k)}}\nonumber\\
\sim{e^{2}p_{F}^{2+\alpha}}
(-1+{c\over {N^{1+\alpha/3}}}+...)
\end{eqnarray}
where $c$ is a positive number
 or, taking into account that $\delta\nu=\nu-\nu_{c}\sim{1\over N}$:
\begin{equation}
E_{1}(\nu)-E_{1}(\nu_{c})\sim {e^{2}p_{F}^{2+\alpha}}|\delta\nu|^{1+\alpha/3}
\end{equation}
We observe that the result (20) retains an explicit dependence
 on the from of the interaction potential, although the latter appears
 to be different form the naive exchange energy (16).
Notice that the cusp has a mirror symmetry about $\nu_{c}=1/2$ required by
a particle-hole symmetry for spin-polarized fermions.

We  stress that the exponent in (20) is greater than one which implies
that states in the vicinity of half filling are closer in energy to
the $\nu={1\over 2}$ state than one could expect on general
grounds (linear cusp). It is also consistent with the fact that the ground
state at $\nu=1/2$ is compressible.

 The present approach also makes it possible
to get an estimate of quasiparticle energy gaps.
In the system with fixed number of particles
the lowest neutral excitation ("exciton")
can be associated with the pole of the charge-density response function
$\chi_{\rho}
(\omega, {\vec k})$ [4]. A spatial separation of a particle and a hole
$(\delta r\sim k/B)$ constituting the exciton icreases with $k$.
At $k\rightarrow \infty$ one obtains  a pair of distant charged
quasihole and quasiparticle with a total energy \\
$\Delta=\omega(k=\infty)=\epsilon_{p}+\epsilon_{h}$.

A quasihole (quasiparticle) excitation
can be viewed as an addition (removal) of $1/N$ flux unit
 of an external magnetic field to the system.
A corresponding  perturbation of incompressible electron fluid
causes a local density distortion which
carries fractional electric charge $\pm {e\over {2N+1}}$
and obeys anyonic statistics $\theta=\pi({2N-1\over{2N+1}})$
as a result
of the fermion liquid polarization. More precisely, each fractionally
charged quasiparticle in the bulk is accompanied by a complementary
charge $\pm e(1-{1\over N})$ located on the boundary.
In the case of a neutral bulk
excitation a total charge on the boundary cancels out.

To estimate a scaling behavior of the gap
function $\Delta(\nu)$ we proceed with a consideration close to that of
the Ref. [8] where dispersions of collective excitations
 were found in the context of the Integer Quantum Hall Effect.
In this case the exciton is merely a transition of a fermion from the
$N^{th}$ Landau level to the $(N+1)^{th}$ one.
In contrast to the case of IQHE one has to start from the bare dispersion
which is completely degenerate $(\omega(k)=0)$. This zero order
approximation is consistent with a proper account of gauge fluctuations
in the limit of  arbitrarily small interaction strength $e\rightarrow0$.

The only contribution  to the exciton dispersion which
remains finite at large $k$
comes
from the exchange self-energy corrections $\Sigma_{N,N+1}$
to the $N^{th}$ and $(N+1)^{th}$
mean field Landau levels. The other terms which can be understood as
a particle-hole binding energy decay as $\delta\omega(k)
\sim -e^2({B\over k})^{\alpha}$.

In the first order approximation the
 exchange self-energy associated with the $n^{th}$ Landau level
is a real constant which can be written
in the form
\begin{equation}
\Sigma_{n}=\int {d\omega d{\vec k}\over {(2\pi)^3}}
\sum_{l=0}^{n}|<n|J_{0}({\vec k})|l>|^2
G_{l}(\omega)<a_{0}(\omega,{\vec k})a_{0}(-\omega,-{\vec k})>
\end{equation}
where $G_{l}(\omega)=(B(n-l)-\omega+i\delta)^{-1}$
and the propagator of the temporal component of the gauge field
$<a_{0}(\omega,{\vec k})a_{0}(-\omega,-{\vec k})>$
can be obtained by inverting (3).

In the case of IQHE considered in [8]
there is a nonzero energy gap $\Delta_{0}={B\over m}$ at $e\rightarrow 0$
and therefore one
may use an unambiguous prescription to evaluate  corrections
as $\delta\Delta = \Sigma_N -\Sigma_{N+1}$.
In our problem a spurious cyclotron frequency disappears under normalizing
the exciton dispersion  $\omega(k)$ with respect
to the case $e=0$ when it becomes
flat. Then one can no longer use the above
 definition for $\Delta$ in terms of $\Sigma_{n}$.
However we consider $\Delta\sim max(\Sigma_N;\Sigma_{N+1})$
as a reasonable estimate
although it may also overestimate the correct result.
Then using (21)  we get  the following relation
\begin{equation}
\Delta(N)\approx \int_{0}^{\infty} {d\omega\over {2\pi}}\int_{0}^{\infty}
 {d{\vec k}\over {(2\pi)^2}}\sum_{l=0}^{N-1}|<N|J_{0}({\vec k})|l>|^2
{G_{l}(\omega)V(k)k^2\over{det{\hat K}_{0}}}
\end{equation}
On neglecting the fermion polarization operator in the denominator of (22)
one obtains a naive result  $\Delta(N)\approx V({\sqrt B})\sim e^2 p_{F}
^{\alpha}/N^{\alpha/2}$ which coincides with the mean field
exchange energy in the absence of gauge fluctuations [9].

To perform a more complete account
of gauge fluctuations we must keep $\Pi_{\mu\nu}(\omega, {\vec k})$
in $det{\hat K}_{0}$. Again using the semiclassical approach [7]
and integrating over continuous variable $\xi$  substituting
$B(N-l)$ we arrive at the integral
\begin{equation}
\Delta(N)\approx \int_{B}^{\infty} {d\omega\over {2\pi}}
\int_{0}^{\infty}
{dk\over
{(2\pi)^2}}
Im{V(k)k^{2}
\over{k^{2}+(4\pi)^2(\kappa
-i\gamma\omega/k)(\chi k^2+i\gamma\omega/k)}}
\end{equation}
Assuming that away from hal filling the quasiparticle spectrum immediately
develops a gap we
subtract from (23) its value at $B=0$
which yields a renormalization
of the chemical potential.
 The resulting quasiparticle gaps obey the scaling law
\begin{equation}
\Delta(\nu)\sim {e^{2}p_{F}^{\alpha}}|\delta\nu|^{{1\over 3}(2+\alpha)}
\end{equation}
 To implement this estimate one has to take into account that the physical
interaction potential $V(k)$
may have different values of $\alpha$ at different scales of $k$.
As a most physically relevant example,
 the bare Coulomb potential has $\alpha =1$
but it changes
to $\alpha =2$ at $k<\kappa_{D}\sim {e^{2}m\over {\varepsilon_0}}$
due to the Debye
screening
( here ${\varepsilon_0}$
is the background dielectric constant).
Thus the Fourier transform of the Coulomb potential
has to be described by different values of $\alpha$ depending on the relation
between relevant momenta $k\sim p_{F}/N$ and $\kappa_{D}$.
Then
one easily obtains that the dependence $\Delta(\nu)\sim |\delta\nu|$ holds
for $\delta\nu >{e^2 m\over {\varepsilon_0 p_{F}}}$ while at smaller
$\delta\nu$ it changes
to $\Delta(\nu)\sim |\delta\nu|^{4/3}$.

 This crossover behavior appears to be in a qualitative agreement
with recent quasiparticle gap measurements [10].
It was found in [10] that
$\Delta(\nu)$ grows linearly with deviation from $\nu_{c}=1/2$
  but an extrapolation of the
linear plot to half
filling yields a negative intercept which can be interpreted as
a sign of an
essentially nonlinear dependence of $\Delta(\nu)$ at small $\delta\nu$.

However, as an alternative explanation
 of these experiments
one could suggest
that the gap simply remains zero in some interval of fractions.
 In fact, our present consideration
 doesn't allow us
to answer the question whether a gap opens at some small
deviation from half filling or whether
it happens only at some finite $\delta\nu_{c}$.
In the latter case we expect that the scaling behavior (24) can only take place
at $\delta\nu_{c}<<\delta\nu<<1$.

The scaling behavior
(24) must be compared with the predictions made in ref.[4]
\begin{equation}
\Delta(\nu)\sim {\rho\over {N m^{\star}}}
\end{equation}
where $m^{\star}$ is an effective mass of a quasiparticle.
 The authors of ref.[4] argued that gauge fluctuations
can be accounted for by including the effective mass renormalization.
This was found in the form $m^{\star}\sim ({\mu\over \Delta})^{1/3}$
in the case of short-range interactions  $(\alpha=2)$
versus $m^{\star}\sim
\ln{\mu\over{\Delta}}$
for the Coulomb interaction $(\alpha=1)$.
Then (25) yields
$\Delta(\nu)\sim {|\delta\nu|^{3/2}}$
for the case of short-range interactions and
$\Delta(\nu)\sim {|\delta\nu|\over {C+\ln|\delta\nu|^{-1}}}$
for the Coulomb interaction.

At the same time it was mentioned in [4] that the available results
for the Coulomb interaction
 at small $N$
obtained from a diagonalization of small systems [11]
can be better fitted by the linear dependence
 $\Delta(\nu)\sim |\delta\nu|$.

In fact, the statements of the ref.[4]  were made on the basis of some
ansatz for the one-particle self-energy caused by infrared divergent
corrections due to the long-range transverse gauge field
$a_{\perp}$.
At $\nu=1/2$  the lowest order contribution was found
 in the form
\begin{equation}
\Sigma_{1}(\epsilon)\approx -g^{2}{p_{F}\over {m\chi^{2/3}\gamma^{1/3}}}
(i\epsilon)^{2/3}\nonumber
\end{equation}
The ansatz proposed in [4] to account for all higher order corrections
is equivalent to the statement that both real and imaginary parts of
the exact $\Sigma(\epsilon)$ are proportional to $\epsilon$
at $\epsilon <<\mu|{p-p_{F}\over p_{F}}
|^{3}$.

On the other hand, a
recent
investigation of
this problem in the framework of the eikonal approximation [12]
which is supposed to be adequate for singular fermion interactions
leads to different conclusions. Namely, it
was found that transverse gauge
fluctuations completely
destroy a pole structure of the bare one-particle Green function.
The asymptotical behavior near the (Luttinger) Fermi surface
found in the eikonal approximation is essentially nonpole-like
\begin{equation}
G(\epsilon, p\approx p_{F})\sim {\mu^{1/4}\over {\epsilon^{5/4}}}
\exp(-{\mu^{1/2}\over {\epsilon^{1/2}}})
\end{equation}
A similar form of the one-particle Green function was also
obtained in [13] by means of a different method.

An additional study shows that due to the intrinsic Ward identities
the effect of infrared
singularities manifested in (27) on longwavelength asymptotics
of  $\nu=1/2$ response functions is
not very prominent.  It also turns out that the exponential singularity
revealed in (27) makes $\nu=1/2$
response functions regular at momenta close to $2p_{F}$
[14]. It would be interesting to find out whether an analogous (partial)
cancellation
between self-energy and vertex corrections
at $\nu={N\over {2N\pm 1}}$
results to
electromagnetic response functions
satisfying the $f$-sum rule and the Kohn theorem whose crucial
importance for a correct description was stressed in [15].

Moreover, on the basis of
 the results of the eikonal approximation an effective free boson description
of the gauge dynamics in $\nu={1\over 2}$ state was constructed [13,14].
An attempt to estimate
temperature $({T\over \epsilon_{F}})$
corrections to the ground state energy $E(\nu={1\over 2}, T)$ performed in the
framework of this effective representation yields a result
consistent with RPA [14].
This fact can be considered as an implicit argument in favor of the
 hypothesis that RPA
is capable to give leading $1/N$-corrections in a weak magnetic
field as well.

In addition, it was shown by Wiegmann and Larkin [7] that the
three-loop "beyond  RPA"
contributions to the ground state energy of 2D fermions
interacting via transverse gauge field
cancel each other up to irrelevant
terms. We consider this fact as another argument in favor
of our conjecture that RPA formulae (4),(22) capture relevant physics.

In conclusion, we study the dependence of the ground
state energy as well as  quasiparticle gaps on filling fraction
in the vicinity of $\nu={1/2}$.
In the lowest order in interaction strength we find
$\delta E(\nu)\sim |\delta\nu|^{1+\alpha/3}$ and
$\Delta(\nu)\sim |\delta\nu|^{{1\over 3}(2+\alpha)}$
where the exponent $\alpha$ describes a power-like decay of the effective
interaction potential $V(r)\sim {1\over r^{\alpha}}$.

The author is indebted to Profs. P.B.Wiegmann, F.D.M.Haldane
 and R.Morf for valuable
discussions of these
and related issues. He is also grateful to Prof. T.M.Rice for
the hospitality
extended to him in ETH-Zurich where this paper was completed.
 This work was supported in part by the US Science and Technology
Center for Superconductivity (Grant NSF-STC-9120000) and by the Swiss National
Fund.
\pagebreak

\end{document}